\input harvmac
\overfullrule=0pt
%

\def\bar{\overline}
\def\x{{\times}}
\def\ra{{\rightarrow}}

\Title{ \vbox{\baselineskip12pt
\hbox{hep-th/0101129}
\hbox{HUB-EP-01/03}}}
{\vbox{\centerline{Fibrewise T-Duality}
\bigskip\centerline{for D-Branes on Elliptic Calabi-Yau}}}
\centerline{Bj\"orn Andreas$^{1}$, Gottfried Curio$^{2}$,
Daniel Hern\'andez Ruip\'erez$^{3}$ and Shing-Tung Yau$^{1}$}
\bigskip
\centerline{\it $^{1}$Department of Mathematics,
Harvard University, Cambridge, MA 02138, USA}
\smallskip
\centerline{\it $^2$Humboldt-Universit\"at zu Berlin,
Institut f\"ur Physik, D-10115 Berlin, Germany}
\smallskip
\centerline{\it $^{3}$Departamento de Matem\'aticas,
Universidad de Salamanca,  37008, Salamanca, Spain}

\baselineskip 14pt
\def\sqr#1#2{{\vbox{\hrule height.#2pt\hbox{\vrule width
.#2pt height#1pt \kern#1pt\vrule width.#2pt}\hrule height.#2pt}}}

\smallskip
\smallskip
\smallskip
\smallskip
\smallskip
\smallskip
\smallskip
\smallskip
\smallskip
\noindent

Fibrewise T-duality (Fourier-Mukai transform)
for D-branes on an elliptic Calabi-Yau
$X$ is shown to require naturally an appropriate twisting of the operation
respectively a twisted charge. 
The fibrewise T-duality is furthermore expressed through known monodromies
in the context of Kontsevich's interpretation of mirror symmetry.

\Date{}

\lref\bada{B. Andreas, G. Curio, D. Hern\'andez Ruip\'erez and S.-T. Yau,
``Fourier-Mukai Transform and Mirror Symmetry for D-Branes and
Elliptic Calabi-Yau'', math. AG/0012196.}

\lref\FMW{R. Friedman, J. Morgan and E. Witten, ``Vector Bundles and F-
Theory,'' Commun. Math. Phys. {\bf 187} (1997) 679, hep-th/9701162.}

\lref\DOPWe{R. Donagi, B. Ovrut, T. Pantev and D. Waldram,
``Standard Models from Heterotic M-theory'',
hep-th/9912208.}

\lref\diacrom{D.-E. Diaconescu and C. R\"omelsberger, ``D-Branes and Bundles
on Elliptic Fibrations'', hep-th/9910172.}

\lref\morr{P. Candelas, A. Font, S. Katz and D. Morrison, ``Mirror Symmetry
for Two Parameter Models- II'', hep-th/9403187.}

\lref\bridge{T. Bridgeland, `` Fourier-Mukai transforms for elliptic
surfaces'', J. reine angew. Math. {\bf 498}, alg-geom/9705002.}

\lref\dave{P. Candelas, X. De La Ossa, A. Font, S. Katz and D. R. Morrison,
``Mirror symmetry
for two parameter models I'', Nucl. Phys. {\bf B416} (1994) 481,
hep-th/9308083.}

\lref\bart{C. Bartocci, U. Bruzzo, D. Hern\'andez Ruip\'erez and J.
M. Mu\~noz  Porras, ``Mirror symmetry on K3 surfaces via
Fourier-Mukai transform'', Commun. Math. Phys. {\bf 195} (1998), 79-93}

\lref\RPo{D. Hern\'andez Ruip\'erez and J. M. Mu\~noz Porras,
``Structure of the  moduli space of stable sheaves on elliptic
fibrations'', math. AG/9809019.}

\lref\vafa{C. Vafa, ``Extending Mirror Conjecture to Calabi-Yau 
with Bundles'', hep-th/9804131.}

\lref\CDo{G. Curio and R. Donagi, ``Moduli in {N=1} Heterotic/F-Theory
 Duality'', Nucl.Phys. {\bf B518} (1998) 603, hep-th/9801057.}

This paper constitutes a shortened and simplified version of \bada\ 
(where details and 
references are given)
intended to present the essential results to an audience of
physicists.

The last years saw an intense study on BPS D-branes, with their
associated bundles, in type II string
theories on a Calabi-Yau manifold.
For the class of elliptic Calabi-Yau's
not only the description of bundles on the whole Calabi-Yau, related to
D6-branes, can be made explicit via spectral
covers (cf. \FMW ), but
it allows also for a version of the Fourier-Mukai (FM)
transform of a bundle $V$
on non-toroidal spaces which in contrast to earlier transforms in such
cases (say on $K3$) sticks completely to the idea of using the duality on
a torus by building a fibrewise FM transform. This 'fibrewise
T-duality' operates on the spectrum of D-branes and has also
a close connection with the spectral cover construction.

We compute the cohomological invariants of a bundle and its
fibrewise dual and
show how {\it the adiabatic character}\foot{I.e., decomposing the
cohomology into base and fibre
parts, the operation of fibrewise duality takes the form
of an adiabatic extension of the same operation on the
(cohomology of a) $T^2$, fulfilling the expectations from the
interpretation as T-duality on D-branes.}{\it of the operation is confirmed
only by using an appropriately twisted charge.}
Alternatively a version of Mukai's {\bf f}-map,
using the usual Mukai vector as charge but a slightly twisted
operation, has to be used.

Also, in a version of the
mirror correspondence, which brings supersymmetric D-branes
on both sides into the play,
bundles, or better sheaves (think of them as bundles supported on
holomorphic subvarieties), on a Calabi-Yau $X$ in type IIA 
are related with special Lagrangian submanifolds 
(with an $U(1)$ bundle over them) in the mirror Calabi-Yau $Y$ with
cohomological invariants relating $H^{even}(X)$ and $H^3(Y)$.
Furthermore, the cohomological effect of 
a line bundle twist (or more general operations related to FM transforms)
is related with a type IIB monodromy
resp. (using the mirror identification) with a type IIA monodromy
(around the locus in moduli space where the divisor vanishes).
{\it The fibrewise FM transform is given its place 
in Kontsevich's general association of FM transforms with monodromies}.

Consider an $SU(n)$ bundle $V$ of $c_1(V)=0$ over an elliptically fibered
$X$ or equivalently $n$ D6-branes wrapped over $X$ with induced
D2-and D0-brane charges (D2i-charges meaning here for now just $ch_i(V)$).
The T-duality on the $T^2$ fibre maps
(subscripts indicate that fibre $F$ or base $B$ is contained
(resp. contains) the wrapped world-volume)
\eqn\Dtual{\eqalign{ D6&\rightarrow \tilde{D4}_B\cr
D4_B\rightarrow \tilde{D6}\;\; &,
\;\; D4_F\rightarrow\tilde{D2}_B \cr
   D2_B\rightarrow \tilde{D4}_{\tilde{F}}\;\; &,
\;\; D2_F\rightarrow  \tilde{D0}\cr
                    D0&\rightarrow \tilde{D2}_{\tilde{F}}}}
The operation of fibrewise T-duality can be described
on the cohomological data representing $V$ and its
FM dual $\tilde{V}$, i.e. we will mirror the mentioned
D-brane relations as a map between $ch(V)$ and
$ch(\tilde{V})$ (this will be made precise). Assume 
the following decomposition
\eqn\decomp{\pmatrix{
H^0(X)&=&{\bf C}\cr
H^2(X)&=&{\bf C}\sigma \oplus H^2(B)\cr
H^4(X)&=&H^2(B)\sigma \oplus {\bf C}\cr
H^6(X)&=&{\bf C}}}
Then essentially the six entries of the Chern character
vector in our decomposition are
pairwise interchanged and the transformation has the block-diagonal form
\eqn\genmatrix{
Q=\pmatrix{
0&*&&&&\cr
{*}&0&&&&\cr
&&0&*&&\cr
&&*&0&&\cr
&&&&0&*\cr
&&&&*&0} \tilde{Q}}
So the cohomological effect of the fibrewise FM
transform (T-duality) is given by adiabatic
extension of the effect of T-duality on the
fibre
where the matrix is \bridge\
\eqn\ellmatrix{
A=\pmatrix{
0&1\cr
-1&0}}
So one would like to see as actual form of the transformation matrix
\eqn\precisematrix{
Q=\pmatrix{0&1&0&0&0&0\cr
-1&0&0&0&0&0\cr
0&0&0&1&0&0\cr
0&0&-1&0&0&0\cr
0&0&0&0&0&1\cr
0&0&0&0&-1&0}\tilde{Q}}
thereby confirming the 'adiabatic' interpretation of the Fourier-Mukai
transform as
(with entries now consisting of
$3\x 3$ blocks corresponding to
$H^*(X)=\sigma H^*(B)\oplus \pi^* H^*(B)$)
\eqn\adiabmatrix{
\pmatrix{
{\bf 0}_3&{\bf 1}_3\cr
-{\bf 1}_3&{\bf 0}_3}}
{\it The Fourier-Mukai transform}

The spectral cover construction\foot{An $SU(n)$ bundle $V$ on $X$ 
of $c_1(V)=0$ decomposes on the typical
fibre $E$ (where $V$ is assumed to be semistable) as a sum
of line
bundles of degree zero each of which
corresponds (a reference point $p$ chosen) to a
point $Q_i$ on $E$. When globalized this leads to a
hypersurface $i: C\hookrightarrow X$, a ramified $n$-fold 'spectral
cover' of $B$, whose equation can still be
twisted so that $C= n\sigma + \eta$ in cohomology. As {\it fibrewise}
$V=p_* {\cal P}$ and 
a twist by a line bundle $L$ over $C$ leaves the fibrewise
isomorphism class unchanged this generalizes to
$V=p_*(p_C^*L\otimes {\cal P})$
where $p$ and $p_C$ are the projections on the first and second factor
of $X\times_B C$.
The condition $c_1(V)=0$ translates to the
fixing $\pi_*(c_1(L))=-\pi_*(c_1(C)-c_1)/2)$
of $c_1(L)$ in $H^{1,1}(C)$ up to a class in $ker \;
\pi_*:H^{1,1}(C)\rightarrow H^{1,1}(B)$.}
describes the bundle $V$ from a line bundle $L$ on the spectral cover
surface $C$.
For the description of the FM transform we will
instead of working on $X\times_B C$ work on $X\times_B \tilde X$
\eqn\mats{\matrix{\;\;\; X\times_B{\tilde X}&\buildrel p_2 \over
\longrightarrow&{\;\;\tilde X}\cr
\scriptstyle{p_1} \biggr \downarrow & &
\scriptstyle{\pi_2}\biggr \downarrow \cr
     \;\;\; X&\buildrel {\pi_1} \over\longrightarrow&\;\;B}}
where $\tilde X$ is the compactified relative Jacobian of $X$. $\tilde
X$ parameterizes torsion-free rank 1 and degree zero sheaves of the
fibres of $X\to B$  and is actually isomorphic with $X$  (see
\bart\ or \RPo). We will identify $\tilde X$ and $X$.
$V$ is constructed from the Poincar\'e sheaf ${\cal P}$ via
\eqn\bundv{V=p_{1*}(p_2^*(i_*L)\otimes {\cal P})}
with
${\cal P}=
{\cal O}(\Delta - \sigma \x \tilde{X} - X\x \tilde{\sigma}- c_1(B))$
normalized to be trivial along $\sigma \x \tilde{X}$
and $X\x \tilde{\sigma}$.

Let us determine the FM-transform.
For this we make use of the fact that the representation of $V$ by the
$(C,L)$ data already looks in itself like a FM transform. So the 
FM transform of $V$ is practically the
double transform of $i_* L$. But the inverse
transform of FM is not precisely FM itself, but rather a slightly
twisted version of the first transform, as we will recall. 
Only this twisted version brings us back from
$V$ to $i_* L$, so if we start on $V$, making
the original FM transform we  get
$i_* L$ times the inverse twist.

So for
$V=p_{1*}(p_2^*(i_*L)\otimes {\cal P})$
the 'fibrewise dual' is given in spectral cover data by
\eqn\dulav{{{\tilde V}=R^1p_{2*}(p_1^*(V)\otimes
 {\cal P}^*)=i_*L\otimes \pi_2^*K_B}}
One wants to show that there exists a matrix $M$ in the block
diagonal form \precisematrix\
relating $V$ to $\tilde{V}$ or $V$ to $i_* L$
(the latter option
is simply the inverse process as $V$ is the dual of $i_* L$).

Before going on let us recall
the FM transform for elliptic fibrations (cf. \bart\ ,
\bridge\ , \RPo).
We define the Fourier-Mukai functors $S^i$, $i=0,1$
by associating with every sheaf $V$ on $X$ the sheaf
$S^i(F)$ on
$X$ ($X$ and $\tilde X$ are identified;
${\cal P}$ the Poincar\'e sheaf on the fibre product)
\eqn\sif{S^i(V)=R^ip_{1*}(p_2^*(V)\otimes {\cal P})}
It can be also described as (cf. \FMW\ )
${\cal P}={\cal I}^*\otimes p_1^*{\cal O}(-\sigma)\otimes
p_2^*{\cal O}(-\sigma)\otimes q^*K_B$
with $q=\pi.p_1=\pi.p_2$ and ${\cal I}={\cal O}(\Delta)^*$
the ideal sheaf of the diagonal
immersion $\delta: X \rightarrow X\times_B X$.

We can also define the inverse Fourier-Mukai functors $\hat{S}^i$, $i=0,1$
\eqn\infm{\hat{S}^i(V)=R^ip_{2*}(p_1^*(V)\otimes {\cal P}^*\otimes
q^*K_B^{-1})}
The relationship between these functors is stated best if we consider
the associated functors between the derived categories of complexes of
coherent sheaves bounded from above.
\eqn\assfd{\eqalign{S&: D^{-}(X)\rightarrow D^{-}(X);\ \ 
S({\cal G})=Rp_{1*}(p^*_2({\cal G})
\otimes {\cal P})\cr
\hat{S}&: D^{-}(X)\rightarrow D^{-}(X);\ \ 
\hat{S}({\cal G})=Rp_{2*}(p^*_1({\cal G})
\otimes {\cal P}^*\otimes q^*K_B^{-1})}}
Note that one obtains an invertibility result
\eqn\fcd{S(\hat{S}({\cal G}))= {\cal G}[-1], \ \ 
\hat{S}({S}({\cal G}))= {\cal G}[-1].}
{\it The K3 case}

Consider first the two-dimensional case of $X=K3$ and assume
the decomposition

\eqn\decompkdre{\pmatrix{
H^0(X)&=&{\bf C}\cr
H^2(X)&=&{\bf C}\sigma \oplus {\bf C}\cr
H^4(X)&=&{\bf C}}}

The class
of the spectral curve $C$ on which $i_*L$ is supported is $C=n\sigma+kF$.

The Chern characters of $i_*L$ can be obtained
using Grothendieck-Riemann-Roch for the embedding $i:C\rightarrow
{\tilde X}$, giving with ${ch(i_*L)Td({\tilde X})=i_*(ch(L)Td(C))}$ that
\eqn\grrd{ch(i_*L)=(0,C,n)}
further $ch(V)$ ($V$ is the only FM transform of $i_*L$) is given by
\eqn\chv{ch(V)=(n,0,-k)}
Let us introduce\foot{The functor $T$ was introduced
\bart\ because its inverse transform in the sense of \fcd\ is the ``natural''
one: $T$ is the FM transform w.r.t. the sheaf $\bar
{\cal P}={\cal P}\otimes q^*{\cal O}(F)$  and its inverse functor $\hat{T}$
is the FM transform w.r.t. the dual sheaf $\bar {\cal P}^*$.
$S$ does not have this property as its inverse transform is not the
FM transform with respect to ${\cal P}^*$, but this
twisted by $q^*K_B^{-1}$. So we just
divided this up in two parts of
$K_B^{-1/2}$, distributed among the original and the
inverse transform.}
the new functor
$T(\cdot)=S (\cdot )\otimes \pi^* K_B^{-1/2}
=S (\cdot )\otimes {\cal O}(F)$ (and similarly $T^i(\cdot)=S^i (\cdot )\otimes
\pi^* K_B^{-1/2} =S^i(\cdot )\otimes {\cal O}(F)$) so that $T^0(i_*
L)=V\otimes {\cal O}(F)$.  We get
\eqn\tfuc{{ch(T^0(i_* L))=ch(V)(1+F)=M\cdot ch(i_*L)}}
where we reach the matrix we wanted
$$
M=\pmatrix{0&1&0&0\cr
-1&0&0&0\cr
0&0&0&1\cr
0&0&-1&0}.$$
{\it The Calabi-Yau threefold case}

Consider a sheaf $V$ and write its Chern character as
$ch(V)=(n,x\sigma+S,\sigma \eta+aF,s)$
($\eta, S\in p_2^*H^2(B)$), then, 
using the decompostion of the cohomology, we find
$$
ch(V)=\pmatrix{n\cr x\cr S\cr \eta \cr a \cr s}\ ,\qquad
ch(\hat{S}(V))=\pmatrix{0\cr -n\cr \eta+{1\over 2}xc_1\cr -{1\over 2} n
c_1-S\cr
s+{1\over 2}\eta c_1\sigma+{1\over12}xc_1^2\sigma\cr
-{1\over 6}n\sigma c_1^2-a-{1\over 2}\sigma c_1
S+x\sigma c_1^2}
$$
If we multiply $ch(\hat{S}(V))$ by the Todd class
$Td(N)=1-{1\over2}c_1+{1\over12}c_1^2$,\foot{We always confuse the
normal bundle $N$ with its pull-back $\pi^* N$ to $X$} we get
$$
ch(\hat{S}(V))\cdot Td(N)=\pmatrix{x \cr
-n \cr \eta \cr -S \cr s-{1\over 12}x\sigma c_1^2 \cr
-a+x\sigma c_1^2}
$$
When $V$ is fibrewise of degree 0 we have $x=0$ and then
\eqn\Mmat{\pmatrix{x \cr
-n \cr \eta \cr -S \cr s \cr
-a}=\pmatrix{0&1&0&0&0&0\cr
-1&0&0&0&0&0\cr
0&0&0&1&0&0\cr
0&0&-1&0&0&0\cr
0&0&0&0&0&1\cr
0&0&0&0&-1&0}\cdot\pmatrix{n\cr x\cr S\cr \eta \cr a \cr s}}
that is
\eqn\MmathS{Td(N)\cdot ch(\hat{S}(V))=M\cdot ch(V)\,.}
So for fibrewise
degree 0, semistable bundles $V$ we have  $\hat{S}^0(V)=0$
and
$\hat{S}^1(V)=i_*L$ and\foot{There is an analogous equation to \MmathS\ 
for the direct FM
transform $S(V)$: if $x=0$ then
$Td(N^{-1})\cdot ch(S(V))=M\cdot ch(V)$.
For fibrewise degree 0 and semistable bundles $V$ we have
$S^0(V)=0$ and with $S^1(V)=i_*\bar{L}$, we have
$
Td(N^{-1})\cdot ch(i_*\bar{L})=Td(N^{-1})\cdot ch(S^1(V))=-M\cdot
ch(V)$.}
$$
Td(N)\cdot ch(i_*L)=Td(N)\cdot ch(\hat{S}^1(V))=-M\cdot ch(V)\,.
$$
The interpretation\foot{In the three-fold case
the sole use of the T-functor from the $K3$-case
(for the map between the Chern classes of the bundle and its dual)
is insufficient to exhibit \precisematrix\ as transformation matrix.
Rather one needs (just as one has in the usual full, not fibrewise,
FM transform actually to take the Mukai vector
with its twist by $\sqrt{Td(X)}$ and not just $ch(V)$) 
the relevant twisted charge.} 
is that here one must 
either twist the operation a little bit and stick to the usual
Mukai-vector or one keeps the
operation and makes the twist in the usual Mukai-vector more 'relative'.
Explicitly, similarly to the usual twist in the Mukai-vector,
here in the fibrewise situation a twist by $\sqrt{Td(N)}$
with the normal bundle $N=(j^*TX)/TB=j^*T_{X/B}=j^*(TX/\pi^* TB)$ plays a role
(where $j:B\hookrightarrow X$).
For a complex of sheaves ${\cal G}$ on $X$
you twist the standard definition of $ch$ to define  a ``charge''
(where $|i+1|=(-1)^{i+1}$)
\eqn\ch{Q({\cal G})=\sum(-1)^ich({\cal G}^i)(\sqrt{Td(N)})^{|i+1|}}
Then
\eqn\chd{Q(\hat{S}(V))=-ch(\hat{S}^1(V))(\sqrt{Td(N)})\;\;\; ,\;\;\;
Q(V)=ch(V)(\sqrt{Td(N)})^{-1}}
and thus
\eqn\mat{Q(\hat{S}(V))=M\cdot Q(V)}
Since $\hat{S}^0(V)=0$ and $\hat{S}^1(V)=i_*L$ we have
$\hat{S}(V)=i_*L[-1]$ as complexes and \chd\ gives
$$
Q(i_*L[-1])=M\cdot Q(V)=M\cdot Q(S(i_*L))
$$
Alternatively the cohomology effect of the FM transform can be
described by a so-called ${\bf f}$-map. For a
fibrewise FM transform we can also 
introduce\foot{It is defined by
$
{\bf f}_r(x)=p_{1*}(p_2^*(x)\cdot Z_r)
$
where $Z_r=\sqrt{p_2^*Td(T_{X/B})}\cdot ch({\cal P})\cdot
\sqrt{p_1^*Td(T_{X/B})}$. Then
$
\sqrt{Td(T_{X/B})}\cdot ch(S(V))={\bf f}_r\big(ch(V)\cdot
\sqrt{Td(T_{X/B})}\ \big)
$.
The effect of ${\bf f}_r$ on the effective charge
given by the Mukai vector
$Q(V)=ch(V)\cdot\sqrt{Td(X)}$ is described by
${\bf f}_r(Q(V))=Q(S(V))$.}
a relative version
${\bf f}_r: H(X)\to H(X)$.
But modifying ${\bf f}_r$ to
\eqn\fmapdef{{\bf f}(x)=p_{1*}(p_2^*(x)\cdot Z)}
with $Z=\sqrt{p_2^*Td(X)}\cdot ch({\cal P})\cdot
\sqrt{p_1^*Td(X)}$, the effective charge of $V$
transforms to
\eqn\fmap{{\bf f}(Q(S(V)))=M\cdot Q(V)}
when $x=0$. If $V$ is also fibrewise semistable so that
$S^0(V)=0$ and
$S^1(V)=i_*L$, then
$$
-{\bf f}(Q(i_*L))=-{\bf f}(Q(S^1(V)))={\bf f}(Q(S(V)))=M\cdot Q(V)
$$
\noindent
{\it Fibrewise T-duality on D-branes at the sheaf level}

Note that we can also describe 
the fibrewise T-duality maps in \Dtual\ at the sheaf level.
Let us consider the skyscraper
sheaf ${\bf C}(x)$ at a point $x$ of $X$. Its FM
transform $S^0({\bf C}(x))$ is a torsion-free rank one sheaf $L_x$ on the
fibre of $X$ over $\pi(x)$ \foot{With the identification $X\simeq\tilde X$
the point
$x$ corresponds precisely to $L_x$ (see \bart\ or \RPo\ )} as we expect
from \Dtual\
and thus $D0\rightarrow D2_F$.
The topological invariants are $n=x=a=S=\eta=0, s=1$ and
\eqn\fmap{ch_i(S^0({\bf C}(x)))=0, \ \ i=0,1,3, \ \ \ \ ch_2(S^0({\bf
C}(x)))=F}

If we start with ${\cal F}={\cal O}_\sigma$; proceeding as in (3.16) of
\RPo\ we have
\eqn\fmsigma{\eqalign{S^0({\cal O}_\sigma)&={\cal O}_X\,,\qquad
S^1({\cal O}_\sigma)=0\cr S^0({\cal O}_X)&=0\,,\quad\qquad
S^1({\cal O}_X)={\cal O}_\sigma\otimes \pi^*K_B}}
Then ${\cal O}_\sigma$ transforms to the structure sheaf of $X$ and ${\cal
O}_X$ transforms to a line bundle on $\sigma$ as we expect from
\Dtual\
since $D4_B\leftrightarrow D6$. We have as before the transformations at the
cohomology level
($n=0,  x=1,  S=0, \eta={1\over 2}c_1, 
a= 0,  s={1\over 6}\sigma c_1^2$)
and get
\eqn\resu{ch_0(S^0({\cal O}_\sigma))=1,\quad ch_i(S^0({\cal
O}_\sigma))=0,\ \ i=1,2,3}
Finally, let us consider a sheaf ${\cal F}$ on $B$; by \fmsigma\ we have
$S^0({\cal O}_\sigma\otimes \pi^*{\cal F})=\pi^*{\cal F}, 
S^1({\cal O}_\sigma\otimes \pi^*{\cal F})=0,
S^0(\pi^*{\cal F})=0\,  S^1(\pi^*{\cal F})={\cal O}_\sigma\otimes
\pi^*{\cal F} \otimes \pi^*K_B$
Then, a sheaf 
${\cal O}_\sigma\otimes \pi^*{\cal F}=j_*{\cal F}$ 
($j: B\ra X$ is the section) supported on a curve
$\tilde C$ in $B$ embedded in $X$ via $j$ transforms to a  sheaf on the
elliptic surface supported on the inverse image of $\tilde C$ in $X$ and
vice versa. So we have relations
\Dtual\ for fibrewise T-duality on D-branes
at the sheaf level (cf. \DOPWe\ )
\eqn\Dtual{\eqalign{D4_B&\rightarrow \tilde{D6}\cr
D2_B&\rightarrow \tilde{D4}_{\tilde{F}}\cr
D0&\rightarrow \tilde{D2}_{\tilde{F}}}}
{\it On the monodromy interpretation of the FM transform}

We finally ask whether the
Fourier-Mukai transform will be related to a monodromy matrix in the
sense of Kontsevich's proposal.
We are interested first in an explicit map between the topological invariants
of the characteristic classes of the Chan-Paton sheaf $V$ and the brane
charges. Consider the BPS charge lattice which can be
identified with the middle cohomology lattice of the mirror manifold
$H^3({\hat X},{\bf Z})$ and consider (we think of two-parameter models)
the central charge associated to the
integral vector ${\bf n}=(n_6, n_4^1, n_4^2, n_0, n_2^1, n_2^2)$ which is
\eqn\zn{Z({\bf n})=n_6\Pi_1+n_4^1\Pi_2+n_4^2\Pi_3+n_0\Pi_4+
n_2^1\Pi_5+n_2^2\Pi_6.}
One the other side, one has in the large volume limit of $X$ the lattice
of microscopic D-brane charges (which is identified with the K-theory
lattice $K(X)$). Here one considers the effective charge $Q$ of a D-brane
state $\eta$ given by the Mukai vector
$Q=
ch(\eta)\sqrt{Td(X)}\in H^{even}(X)$
with the associated central charge
\eqn\zt{Z(t)=\int_Z {t^3\over 6}Q^0-{t^2\over 2}Q^2+tQ^4-Q^6.}
where $t$ is the generic K\"ahler class and one gets
\eqn\zent{Z(Q)={r\over 6}t^3-{1\over 2}ch_1(V)t^2+(ch_2(V)+{r\over 24}c_2(X))t
-(ch_3(V)+{1\over 24}ch_1(V)ch_2(V))}

The comparison of $Z({\bf n})$ and $Z(t)$ leads then to a map between the low
energy charges ${\bf n}$ and the topological invariants of the
K-theory class $\eta$.

The monodromy transformations $S_D$
correspond to automorphisms $M(D)$
\eqn\shift{ [V]\rightarrow [V\otimes {\cal O}_X(L)], \ \ \ \ \
 [V]\rightarrow [V\otimes {\cal O}_X(H)]}
and we see that the linear transformations acting on ${\bf n}$ corresponding
to $D$ are
\eqn\lintr{M(D)=S_D^{-1}}

Let us consider now a second type of monodromy transformation  proposed by
Kontsevich. He proposed that the monodromy $T$ about the conifold locus
of the mirror corresponds to the automorphism of the derived category
whose effect on cohomology is
\eqn\smat{{\cal S}:\gamma\rightarrow \gamma-\Big(\int \gamma\wedge Td(T_X)\Big)
\cdot {\bf 1}_X}
(${\bf 1}_X$ the standard generator of $H^0(X,{\bf Q})$)
leading to a change in the topological invariants
\eqn\topv{ch(V)\rightarrow ch(V)-({{ch_1(V)c_2(X)}\over 12}+ch_3(V))}
Using the expression of the prepotential in the large radius limit
${\cal F}={1\over 6}(t\cdot J)^3-{c_2(X)\over 24}(t\cdot J)+....$
where $(t\cdot J)=\sum t_aJ^a$ and
using the period vector $\Pi$ we find the expression valid for both models
${\bf P}_{1,1,2,2,2}^4[8]$ and ${\bf P}_{1,1,2,2,6}^4[12]$
\eqn\modeli{\eqalign{ch_1(V)&=n_4^1(J_1-2J_2)+n_4^2J_2\cr
                     ch_3(V)&=-(n_4^1(J_1-2J_2)+n_4^2J_2){c_2(X)\over 12}-n_0}}
leading to the universal shift
\eqn\shiuft{n_6\rightarrow n_6+n_0}
comparing this to the monodromy we find for the linear transformation
acting on ${\bf n}$
\eqn\lina{{\cal S}=T^{-1}}
We now ask whether the
Fourier-Mukai transform will be related to a monodromy matrix in the
sense of Kontsevich's proposal. As the corresponding matrices are by
mirror symmetry identifiable already on the bundle side this comes
down essentially to
the question whether the matrix $M$ in \precisematrix\ ,
is generated by $S_H, S_L, T$.
Recall that the transformations in the Kontsevich association
can themselves be considered as FM
transforms. This time
we will not work on the fibre product but rather on the ordinary product.
\eqn\mats{\matrix{\;\;\; X\times {\tilde X}&\buildrel q_2 \over
\longrightarrow&{\;\;\tilde X}\cr
\scriptstyle{q_1} \biggr \downarrow & &
\scriptstyle{\pi_2}\biggr \downarrow \cr
     \;\;\; X&\buildrel {\pi_1} \over\longrightarrow&\;\;B}}
The Fourier-Mukai functors $S^i$ are
(${\cal E}\in D(X \times X)$ in the derived category)
\eqn\sif{S_{\cal E}^i(V)=R^ip_{1*}(p_2^*(V)\otimes {\cal E})}
We can also define the full FM transformation at the derived
category level $S_{\cal E}({\cal G})=Rp_{1*}(p_2^*({\cal G})\otimes {\cal E})$.
For example the twist transformation
${\cal G}\ra {\cal G}\otimes {\cal L}$ considered in \shift\ comes
from ${\cal E}= {\cal O}_{\Delta}\otimes q_2^*({\cal L})$ where $\Delta$ is the
diagonal of $X\x X$. The other
operation (corresponding to the 'conifold monodromy') considered before
corresponds to an
${\cal E}$ whose cohomology is the ideal sheaf
${\cal I}_{\Delta}$ of the diagonal of $X\x X$. This recovers 
the ``gamma shift'' \smat\ 
\eqn\chKK{ch (S_{{\cal I}_\Delta}({\cal G}))=ch_0(S_{{\cal O}_{X\x X}}({\cal
G}))-ch  ({\cal G}) =\Big(\int ch({\cal G})\wedge Td(T_X)\Big)-ch({\cal G})}

Our fibrewise FM transform is specified by using
${\cal E}=j_*{\cal P}$ where $j: X\times _B X \ra X \x X$ is the natural
embedding\foot{FM on $X\times_B X$ with respect to $P$
is FM on $X\times X$ with respect to $j_*P$}, 
so it is build up out of the divisors
(resp. their associated line bundles) $\sigma$ and
$\pi^*K_B$ on the one hand and by ${\cal O}_{\Delta}$ on the other.

We have the $M$ matrix equation for 
the inverse FM transform. This is ${\hat S}=S_{j_*{\cal Q}}$ with
$ {\cal Q}={\cal P}^*\otimes q^*(K_B^{-1})={\cal I}
\otimes p_1^*{\cal O}(\sigma)\otimes p_2^*{\cal O}(\sigma)\otimes q^*K_B^{-2}
$
where ${\cal I}$ is the ideal of the diagonal of $X\times_B X$. Then
${\hat S}=(\otimes q^*K_B^{-2})\circ(\otimes{\cal O}(\sigma))\circ
S_{j_*{\cal I}}\circ (\otimes {\cal O}(\sigma))$
or
$$
{\hat S}=S_{{\cal O}_\Delta(2c_1)}\circ S_{{\cal O}_\Delta(\sigma)}\circ
S_{j_*{\cal I}}\circ S_{{\cal O}_\Delta(\sigma)}
$$
Having written our fibrewise FM transform ${\hat S}$ as a
composition of three divisor FM transforms and one FM transform
$S_{j_*{\cal I}}$
we expect the following relation for its monodromy, expressed as a product
of divisor monodromies and the monodromy around the diagonal
of the fibre product of ideal sheaf ${\cal I}$
\eqn\monko{S_V=S_{\sigma}\cdot S_{c_1}^{2}\cdot S_{\cal I}\cdot S_{\sigma}}
Let us make this more explicit by considering the model
given by the elliptic fibration
${\bf P}^4_{1,1,1,6,9}[18]$, extensively studied in the context of mirror
symmetry \morr\ and in the context of
D-branes on elliptic Calabi-Yau \diacrom. Among the degree
18 hypersurfaces is\foot{At $z_1=z_2=z_3=0$ the ambient space 
has a singular line which intersects $X$
in a single point. The blow up of this line gives an exceptional
divisor $E=P^2$ in $X$. A second divisor $L$ (defined by the first order
polynomials) is given by the elliptic surface over a line in $P^2$
and together with $E$ generates $H_4(X,{\bf Z})$. The elliptic
fibration structure is induced by the linear system $|L|$ generated by
$z_1, z_2, z_3$ mapping $X$ to ${\bf P}^2$. The section of the fibration
is given by $B_2=E$. The homology class of the elliptic
fibre in $H_2(X)$ will be
denoted by $h=L^2$. Further intersection relations are given by
$E\cdot L^2=1, \ \ E^2\cdot L=-3, \ \ L^3=0, \ \ E^3=9$.
Working in the $E,L$ basis
(cf. \diacrom ) the generic K\"ahler class is given
by $J=t_1E+t_2L$ with $t_1,t_2$ coordinates on the K\"ahler moduli space.
In the degree 18 model again one has $\sigma =E=B=P^2$ and
$\pi^*K_B=3L$; note that $H=3L+E$ with a corresponding multiplicative
relation $S_E=S_H\cdot S_L^{-3}$ for the matrices.}
\eqn\dega{z_1^{18}+z_2^{18}+z_3^{18}+z_4^3+z_5^2=0}
Here the relation becomes
\eqn\rel{S_V=S_E\cdot S_L^6\cdot S_{\cal I}\cdot S_E}
where the commuting (cf. \dave , p. 12) matrices $S_L, S_E$ are given by
%
\eqn\matri{\eqalign{S_L&=\pmatrix{1&0&-1&3&2&0\cr
               0&1&0&1&0&-1\cr
               0&0&1&0&-1&0\cr
               0&0&0&1&0&0\cr
               0&0&0&0&1&0\cr
               0&0&0&1&0&1}, \ \ \
S_E=\pmatrix{1&-1&0&1&0&0\cr
               0&1&0&-9&0&3\cr
               0&0&1&3&0&-1\cr
               0&0&0&1&0&0\cr
               0&0&0&1&1&0\cr
               0&0&0&-3&0&1}}}

In order to obtain the matrices $S_{\cal I}$ and $S_{V}$
we have to use the comparison
(performed for this model in \diacrom)
of the central charges $Z({\bf n})$ and $Z(Q)$
which gives the relation between the middle cohomology charges ${\bf n}$
and the cohomological invariants of the vector bundle ${\cal G}$
\eqn\cchv{{ch({\cal G})=(n\; ,\; \alpha E +\beta L\; ,\; 
\gamma EL+\delta L^2\; ,\; \epsilon) }}
where the coefficients are given in terms of the BPS charge vector {\bf n}
\eqn\coeff{{(n,\alpha,\beta,\gamma,\delta,\epsilon)=
(n_6\; ,\; n_4^1\; , \; n_4^2\; , \; 
{3\over 2}n_4^1+n_2^2\; , \; {3\over 2}n_4^2+n_2^1\; , \;
-n_0+{1\over 2}n_4^1-3n_4^2)}}
The FM transforms ${\hat S}(.)$ and $S_{\cal I}(.)$ induce a linear
transformation on the BPS charge lattice
$$S_{\cal I}=\pmatrix{1&0&3&-9&0&0\cr
           1&1&3&-9&-1&0\cr
           0&0&1&0&0&0\cr
           0&0&0&1&1&0\cr
           0&0&0&0&1&0\cr
           0&0&1&-3&0&1}, \ \ \
S_{V}=\pmatrix{0&-1&0&1&0&3\cr
           1&0&0&0&-1&0\cr
           0&0&0&0&-3&-1\cr
           0&0&0&0&-1&0\cr
           0&0&0&1&0&0\cr
           0&0&1&-3&-3&0}$$
with $S_{\cal I}^{-1}=P(S_{\cal I}(\cdot )), 
S_{V}^{-1}=P({\hat S}(\cdot ))$
denoting the linear transformations on the lattice ${\bf n}$.

Finally, using these matrices we can write the $M$ matrix as
(for $\alpha=n_4^1=0$; $l$ relates (cf. \coeff )
the period basis  with the 'fibration' basis
\decomp\ , $S_{td}$ represents the $Td(N)$ twist )
\eqn\mata{M=l\cdot [S_{V}^{-1}]^t\cdot l^{-1}\cdot S_{td}}
%
%
%

Let us close with some remarks:

1) First we want to point to some analogues on the mirror side.
Note that \mata\ also shows that if we perform a linear
transformation on $H^3(Y)$ (transforming in the ``fibre base'')
by $l$ the $M$ matrix naturally operates on the charge lattice as
$$\pmatrix{n \cr \alpha \cr \beta \cr \gamma \cr \delta \cr \epsilon}=M\cdot
\pmatrix{-\alpha\cr n\cr -\gamma \cr \beta \cr -\epsilon \cr \delta}$$
and thus \mata\ gives just the right {\it transport} of the $M$ matrix
to $H^3(Y)$
$${\matrix{\;\;\; H^{even}(X) &\buildrel M\over
\longrightarrow& H^{even}(X)\cr
\biggr \downarrow & &
     \biggr \downarrow \cr
     \; H^3(Y)&\buildrel {M} \over\longrightarrow&\;\; H^3(Y)}}$$
Thus the $l$ transformation
shows how to get the 'fibration' basis \decomp\
from the K\"ahler period vector
basis, resp. (after identification with the mirror side) how
the \decomp\ basis (when transported via identification to the mirror
side) is related to the complex structure period vector basis,
making explicit the decomposition
in $H^3(Y)$ corresponding to \decomp\
$${H^3(Y)=H^{non-ell}\oplus H^{ell}}$$
One can ask whether this decomposition
and 'duality' transformation on the middle {\it cohomology} of $Y$
comes actually from a natural decomposition and map on the 
{\it space} $Y$. Also, in the context of the extended mirror conjecture
\vafa\ with its $h^1(End(V),X)=h^1(C)$, one can
transport detailed information about $V$ from 
index computations.

2) For the line bundle $L$ on the spectral cover $C$ one has
$$
0\rightarrow H^1(C,{\cal O}_C)
          \rightarrow Ext^1_{X}(i_*L,i_*L)\rightarrow H^0(C,{\cal O}_C)
\rightarrow 0
$$
so that
\eqn\dimext{\eqalign{ {\rm dim} \ Ext^1_D(L,L)&=h^1(C,{\cal O}_D)
=h^{(0,1)}(C)\cr
{\rm dim} \ Ext^1_{X}(i_*L,i_*L)&=h^1(C,{\cal O}_D)
+h^0(C,N)=h^{(0,1)}(C)+h^{(2,0)}(C)}}
The vector bundle $V=S^0(i_*L)$ has as
unique inverse FM transform
${\hat S}^1(V)=i_*L$,so
\eqn\parc{Ext^1_X(V,V)=Ext^1_X({\hat S}^1(V),{\hat S}^1(V))
=Ext^1_X(i_*L,i_*L)}
(by the ``Parceval isomorphism'') and  then
\eqn\dimextv{\eqalign{ {\rm dim} \ Ext^1_D(L,L)&=h^1(C,{\cal
O}_D)=h^{(0,1)}(C)\cr {\rm dim} \ Ext^1_{X}(V,V)&=h^1(C,{\cal O}_D)
+h^0(C,N)=h^{(0,1)}(C)+h^{(2,0)}(C)}}
Actually one can see {\it explicitly}
that ${\rm dim} Ext^1_X(V,V)={\rm dim} Ext^1_X(i_*L,i_*L)$ without using
\parc\ : use the character valued index and compute in the spectral
cover representation (cf. \CDo\ ) the relation
${\rm dim}\; H^1(X,End(V))=h^{(2,0)}(C)+h^{(1,0)}(C)$
showing the  agreement.

\listrefs
\bye